\documentclass[prl,aps,twocolumn,showpacs]{revtex4-1}
\usepackage[centertags]{amsmath}
\usepackage{graphicx}
\usepackage{amsfonts}
\usepackage{epstopdf}
\usepackage{color}
\usepackage[dvipsnames,svgnames]{xcolor} 

\begin{document}

\title{``Extraordinary'' modulation instability in optics and hydrodynamics}

\author{Guillaume Vanderhaegen$^1$, Corentin Naveau$^1$, Pascal Szriftgiser$^1$, Alexandre Kudlinski$^1$, Matteo Conforti$^1$, Arnaud Mussot$^{1,2}$, Miguel Onorato$^3$, Stefano Trillo$^4$, Amin Chabchoub$^{5,6}$ and Nail Akhmediev$^7$} 

\affiliation{$^1$University of Lille, CNRS, UMR 8523-PhLAM-Physique des Lasers Atomes et Mol\'ecules, Lille, France}
\affiliation{$^2$Institut Universitaire de France (IUF), France}
\affiliation{$^3$Dipartimento di Fisica, Università degli Studi di Torino, 10125 Torino, Italy}
\affiliation{$^4$Department of Engineering, University of Ferrara, 44122 Ferrara, Italy}
\affiliation{$^5$Centre for Wind, Waves and Water, School of Civil Engineering, The University of Sydney, Sydney, NSW 2006, Australia} 
\affiliation{$^6$Marine Science Institute, The University of Sydney, Sydney, NSW 2006, Australia}
\affiliation{$^7$Department of Theoretical Physics, Research School of Physics, The Australian National University, Canberra, ACT 2600, Australia}

\begin{abstract}The classical theory of modulation instability (MI) attributed to 
Bespalov-Talanov in optics and Benjamin-Feir for water waves is just a linear approximation of nonlinear effects and has limitations that have been corrected using the exact weakly nonlinear theory of wave propagation. We report results of experiments in both, optics and hydrodynamics, which are in excellent agreement with nonlinear theory. These observations clearly demonstrate that MI has wider band of unstable frequencies than predicted by the linear stability analysis. The range of areas where the nonlinear theory of MI can be applied is actually much larger than considered here.
\end{abstract} 

\maketitle 

\section{Introduction}

Well-known Bespalov-Talanov (BT) \cite{Bespalov} and Benjamin-Feir (BF) 
  \cite{benjamin1967disintegration,benjamin1967rspa} 
instabilities discovered more than 60 years ago (1966 and 1967, respectively) played a significant role in understanding nonlinear phenomena in optics and hydrodynamics   \cite{lake1977nonlinear,yuen1982nonlinear,tulin1999laboratory,kimmoun2016modulation,tai1986exp,mussot2014fermi,Mussot2018,pierangeli2018observation}.
The detailed description of these fundamental results can be found in any common book in nonlinear optics \cite{remoissenet2013waves,Agrawal}, ocean waves  \cite{mei1983applied,osborne2010nonlinear}, and more generally, in nonlinear dynamics book literature \cite{Daniel}. The  theory tells us that a plane wave or a constant amplitude wave (CW) is unstable relative to small amplitude perturbations with frequencies within certain deterministic and finite range. These perturbations are unstable and can grow exponentially, thus, leading to modulated waves with infinitely high amplitude. Clearly, such growth is unphysical and has to be reconsidered using an approach beyond linear theory. 

Indeed, the accurate nonlinear theory \cite{akhmediev1986modulation} predicts saturation and the maximal amplitude of periodic waves excited due to modulation instability (MI). This prediction is in accordance with conventional wisdom: ``what goes up must come down". In fact, this nonlinear stage of MI predicted not only the exponential growth but the following exponential decay back to the constant amplitude wave \cite{akhmediev1986modulation}. The latter was not obvious and required many years before this seemingly simple principle ``must come down" has been confirmed, first with the observation of growth saturation in water waves \cite{lake1977nonlinear}, and then with the demonstration of the full recursive behavior in optical experiments \cite{Simaeys,vanSimaeys2002josab}. If translated to the frequency domain, this principle is, essentially, the Fermi-Pasta-Ulam recurrence \cite{fpubook} (see also \cite{lake1977nonlinear,yuen1982nonlinear,Heatley,mussot2014fermi,Mussot2018,pierangeli2018observation}) .

Despite these achievements, it has been recently demonstrated that not all secrets of modulation instability concealed by the linear approach have been revealed so far \cite{PRA2020}. The results obtained in \cite{PRA2020} demonstrate that the linear theory does not accurately predict the range of unstable frequencies. This fact is, once again, not obvious. An exact nonlinear theory is essential for revealing the full range of frequencies that are unstable due to the modulation. 
Exact solutions of the nonlinear Schr\"odinger equation (NLSE) that describe the nonlinear stage of modulation instability are presently known as Akhmediev breathers (AB)  \cite{Mahnke,Priya2,Varlot,hammani2011spectral,Bendahmane2015exp,Demetri,Bertrand}. 
The latter form a family of solutions with a free parameter that is directly related to the whole interval of unstable frequencies in the BT and BF theories. However, even the AB solutions do not cover the whole range of unstable frequencies. The family of ABs is actually a particular case of more general family of solutions that have been found in \cite{akhmediev1987exact} and refined recently in \cite{PRA2020}. This extension expands the range of unstable frequencies predicted in the the BT and BF theories. It has important ramifications for theory, experiment and applications \cite{dudley2019rogue}. It means, that periodic perturbations of a plane wave (or CW) can grow in the situations when we would not expect them to do so.

Presenting simultaneously optical and hydrodynamic experiments confirming this exceptional feature of modulation instability in a single work has far reaching consequences. 
Observing the same effect at nearly opposite ends of spatial and time scales of MI in physics is a convincing argument confirming the validity of the new finding. It means that similar phenomena at other scales such as MI in plasma \cite{Ghosh} or in Bose-Einstein condensate \cite{Xiao-Xun} must also be re-examined.
In optics, the extension of the range of frequencies leading to MI might have multiple applications for generating frequency combs \cite{Cundiff2005}, periodic pulse trains \cite{Greer1989} and supercontinuum radiation \cite{dudley2010supercontinuum}. In hydrodynamics, the new findings might result in reconsidering conditions leading to formation of rogue waves in the ocean \cite{kharif2009rogue}.

\section{Theoretical background}

We start with the NLSE written in the normalised form:
\begin{equation}\label{nlse}
i\psi_z+\frac{1}{2}\psi_{tt}+|\psi|^2\psi=0
\end{equation}
where $\psi$ is the wave envelope function, $z$ is the longitudinal co-ordinate, and $t$ is the time in a frame moving with group velocity.
We are interested in doubly periodic waves, e.g., in solutions of Eq.(\ref{nlse}) that are periodic both in space and in time \cite{PRA2020}. They comprise the three-parameter family 
of solutions with a single period along each axis, $z$ and $t$. This family contains as particular cases other `elementary' solutions and families \cite{PRA2020}. To be specific, doubly periodic solutions of Eq.~(\ref{nlse}) can be presented in general form:
 \begin{equation} \label{ansatz}
\psi(t,z)=[Q(t,z) + i\delta(z)]e^{i\phi(z)},
\end{equation} with the functions $Q(t,z)$, $\delta(z)$ and $\phi(z)$ that can be found by a direct substitution of (\ref{ansatz}) into (\ref{nlse}) \cite{akhmediev1987exact}.
There are two forms of such solutions, classified as  A- and B-types depending on the parameters of the family.
Each type contains MI as the limiting case. However, the limiting case of B-type solutions is the standard MI while the limiting case of A-type solutions is more general.

This apparently puzzling asymmetry between the two families finds its physical justification in the fact that A-type solutions can be considered as the full nonlinear dressing of solutions of the NLSE obtained in the linear limit (when dispersion dominates over nonlinearity). As discussed in more details in \cite{PRA2020}, for very high modulation frequencies, the deformation introduced by the nonlinearity is small and essentially the modulation experiences, upon evolution, only a periodic phase shift  \cite{Trillo1991b}. However, when the frequency is reduced to sufficiently small values, the deformation due to the nonlinearity becomes strong, thereby inducing a net amplification of the input sidebands even outside the conventional MI bandwidth. Conversely, B-type solutions start to appear only at frequencies below the conventional band-edge of MI, as a result of the symmetry breaking nature of the onset of conventional MI \cite{Mussot2018,PRA2020}. Therefore, B-type solutions cannot be responsible for any unconventional MI.

Thus, our point of interest in this work is the A-type solutions. 
Then, the three functions in (\ref{ansatz}) are defined as follows.
 Namely, for the function  $\delta(z)$, we have the following expression:
\begin{equation}
\delta(z)=\sqrt{\frac{\alpha_3}{2}(1-\nu)}\sqrt{\frac{1+\mbox{dn}(\mu z,k)}{1+\nu \mbox{cn}(\mu z,k)}}\mbox{sn}(\mu z/2,k),
\end{equation}
where 
$m=k^2=\displaystyle \frac{1}{2}\left(1-\frac{\eta^2+\rho(\rho-\alpha_3)}{AB}\right),$ 
$A^2=(\alpha_3-\rho)^2+\eta^2,$ 
$B^2=\rho^2+\eta^2,$ 
$\nu=\displaystyle\frac{A-B}{A+B},$
and $\mu=4\sqrt{AB}$.
The function $\delta$ varies within the interval $0<\delta^2<\alpha_3$.

The phase $\phi(z)$ is given by:
\begin{eqnarray} \label{psiB}
\nonumber
\phi(z)=\left(2\rho+\frac{\alpha_3}{\nu}\right)z-&
\frac{\alpha_3}{\nu\mu}\bigg[\Pi(\mbox{am}(\mu z,k),n,k)- \\
&-\nu\sigma\tan^{-1}\left(\frac{\mbox{sd}(\mu z,k)}{\sigma} \right) \bigg]
\end{eqnarray}
where
$n= \frac{\nu^2}{\nu^2-1}$, 
$\sigma= \sqrt{\frac{1-\nu^2}{k^2+(1-k^2)\nu^2}}$, and
$\mbox{sd}(\mu z,k)= \frac{\mbox{sn}(\mu z,k)}{\mbox{dn}(\mu z,k)}$, $\Pi(\mbox{am}(\mu z,k),n,k)$ is the incomplete elliptic integral of the third kind with the argument am$(u,k)$ being the amplitude function.

In contrast to $\delta$ and $\phi$, the function $Q$ depends on two variables $t$ and $z$. It is given by:
\begin{equation} \label{QtypeB}
Q(t,z)=sb-c_+\frac{r+\mbox{cn}(pt,k_q)}{1+r\mbox{cn}(pt,k_q)},
\end{equation}
where 
$s(z)= \mbox{sign} \left[ \mbox{cn}(\mu z/2,k) \right] $,
$r= \frac{M-N}{M+N}$,
$p=\sqrt{MN}=2\displaystyle\sqrt[4]{(\alpha_3-\rho)^2+\eta^2},$
$k_q^2=\displaystyle \frac 1 2+2\frac{\rho-\alpha_3}{p^2},$
$b=\sqrt{\alpha_3-y},$  $y(z)=\delta^2(z),$
$c_\pm=\sqrt{2\left[\sqrt{(y-\rho)^2+\eta^2}\pm(\rho-y)\right]},$
$M^2=(2sb+c_+)^2+c_-^2,$ and $N^2=(2sb-c_+)^2+c_-^2.$

These functions and, consequently, the whole family of solutions, depend on three arbitrary real parameters $\alpha_3,\rho,\eta$ \cite{PRA2020,akhmediev1987exact}. 
The periods in $z$ and $t$ also depend on these parameters. They are given by:
$\mathcal{Z}={8K(k)}/{\mu},$ $\mathcal{T}={4K(k_q)}/{p}$, respectively, where $K(k)$  is the complete elliptic integral of the first kind. These free parameters provide us with the possibility of accurately controlling the wave evolution with periodic initial conditions and particularly the development of modulation instability.

\section{Instability outside the conventional MI band}

Equations above provide an exact wave dynamics with two frequencies. 
Thus, the MI which is periodic along the $t$-axis is a particular case of these solutions. Indeed, there is a range of parameters $\rho$ and $\eta$ when the solution represents the growth of a periodic perturbation on top of a continuous wave. This happens when $0<\rho<1$ and $\eta \rightarrow 0$. This range corresponds to exact conditions of modulation instability with the exponential growth of periodic perturbation with a frequency defined by $\rho$. On the other hand, for parameters $\rho$ and $\eta$ beyond this range, the evolution has all features of modulation instability but the growth of the perturbation takes different form. 

\begin{figure}[ht]
\includegraphics[scale=0.47]{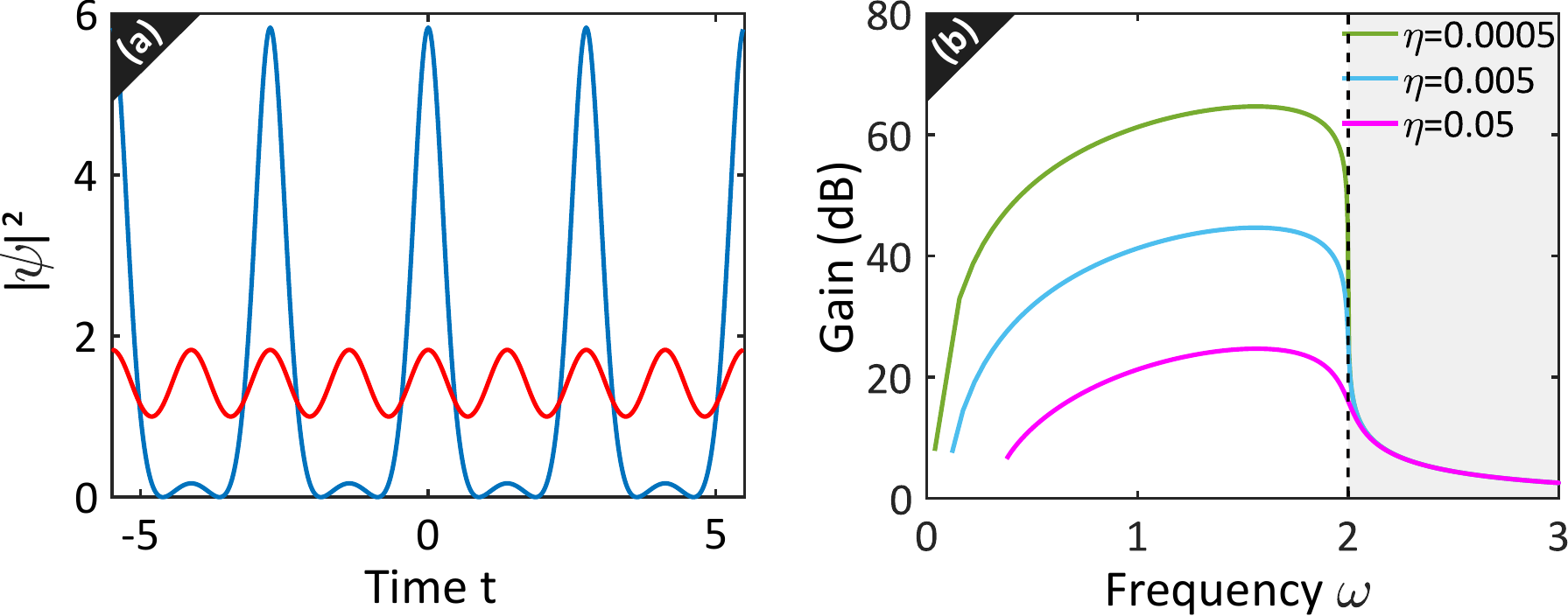}
\caption{(a) Transformation of a periodic perturbation on top of the CW (red curve) into a train of pulses (blue curve). Parameters of the solution here are: $\rho=0$, $\eta=1$, $\alpha_3=1$. Modulation frequency $\omega=2.287>2$ is outside of the instability band. (b) Amplification of periodic component of the solution vs frequency. Frequency $\omega$ depends on the parameter $\rho$ that changes in the interval [$-3,1$]. Grey area beyond dashed black line marks the frequency range $\omega >2$, 
located outside the conventional MI gain band ($\omega \le 2$).
}
\label{growth}
\end{figure}

This more general evolution is periodic in $z$.
The solution is closest to the continuous wave when the evolution variable $z=\pm \mathcal{Z}/4$. Starting from one of these values of $z$ leads to the growth of modulations on the background CW. One example is given in Fig.~\ref{growth}(a) that shows wave intensity profiles at $z=-\mathcal{Z}/4$ when the modulation is small (red curve) and at $z=0$ when the modulation is maximal (blue curve). Pulses within this periodic pulse train are maximally compressed. The wave intensity profile returns back to the initial condition at $z=+ \mathcal{Z}/4$.

The amplification of the periodic component of the solution calculated numerically from the exact solution is shown in Fig.~\ref{growth}(b). Here, the frequency range [$0,2$] is the standard band of modulation instability. Amplification within this range is not surprising. However, the amplification is not zero when the frequency $\omega >2$. The amplification here might seem smaller than within the band [0,2]. However, the amplitudes of the pulse trains reached due to the growth are of the same order of magnitude as within the band.  Thus, the effect is easily measurable in experiments. 
Moreover, the frequency range shown in Fig.~\ref{growth}(b) is nearly 1.5 times the  conventional MI bandwidth $\omega \in [0,2]$. In reality, it is much wider than in this figure. Even from this point of view, the effect is easily observable. 
As can be seen from Fig.~\ref{growth}(b), the value of amplification depends on the parameter $\eta$. For larger values of $\eta$, the amplification within the standard MI band is smaller. However, the amplification outside of this band does not depend on $\eta$. Thus, at larger values of $\eta$, the MI effect is nearly the same order of magnitude within and outside of the standard band.

Another remarkable feature of the MI visible in Fig.\ref{growth}(a) is the period of the pulse train which is twice the period of the initial modulation. Every second maxima of the periodic perturbation grows while the juxtaposing maxima decay. This feature adds flexibility to potential applications of the effect. The red curve in Fig.~\ref{growth}(a) and analogous curves calculated for other values of parameters have been used as initial conditions in the optical and water wave experiments as well as in numerical simulations presented below.

\section{Optical experiments}

\begin{figure}[ht]
\includegraphics[scale=0.25]{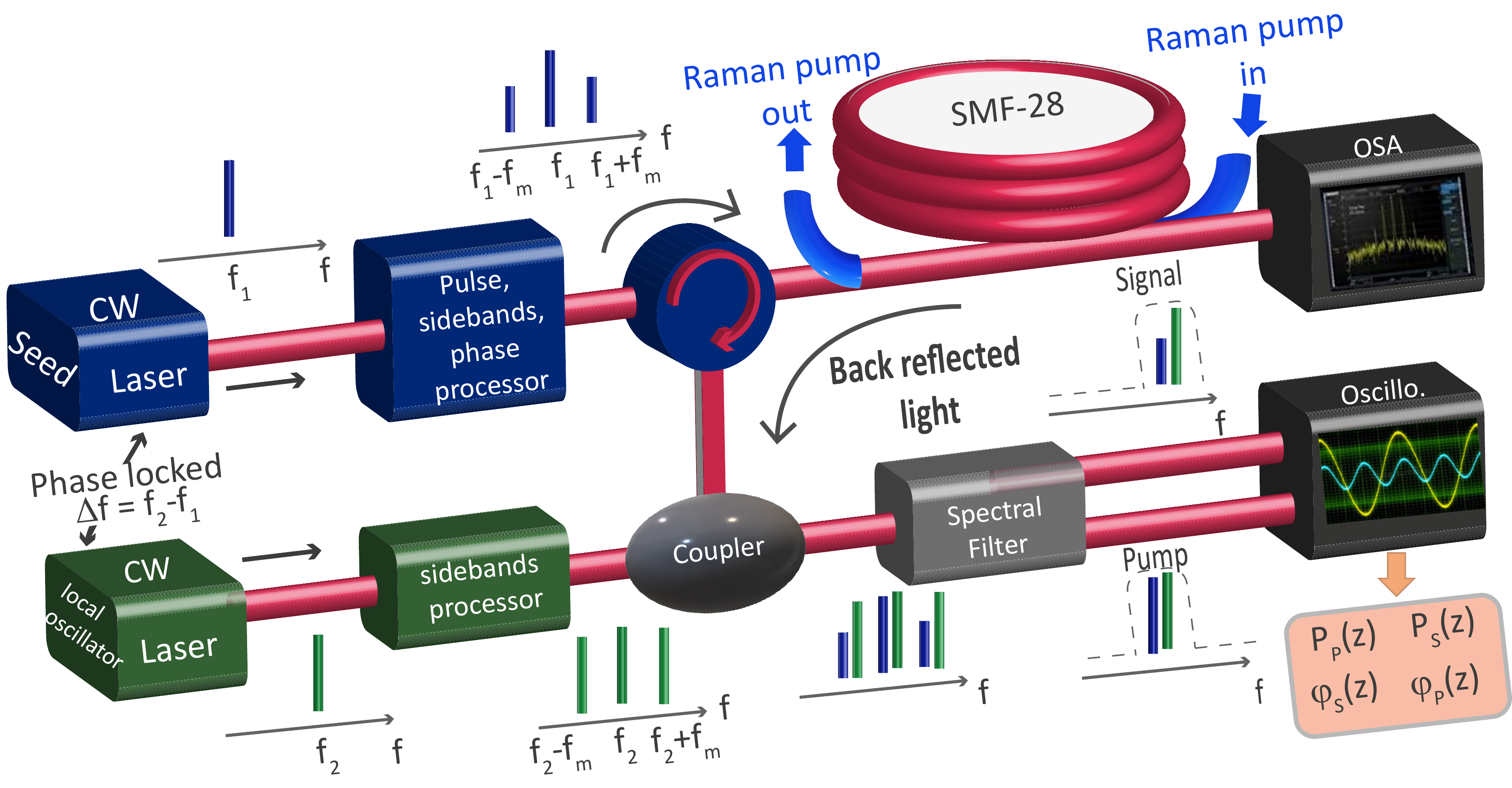}
\caption{Experimental setup: $f_{1,2}$ are the frequencies of the main laser and the local oscillator laser, respectively. Here  $f_{m}$ is the input modulation frequency (pump frequency at $f_1$, input sideband frequencies at $f_1 \pm f_m$). The backscattered signal from the SMF-28 fiber goes through a circulator to be analysed via heterodyning (beating with the local oscillator) and then filtered (waveshaper) to isolate the power and phase evolutions of the pump and the first-order side-band pair in the MI spectral comb; OSA - optical spectrum analyzer.}
\label{Fig1_Exp}
\end{figure}

For optical experiment, we used a setup similar to the one used in 
 \cite{Mussot2018,Vanderhaegen2020exp} and devoted to investigate nonlinear stage of MI within its conventional bandwidth. Its schematic is shown in Fig.~\ref{Fig1_Exp}. 
The input in the form of continuous wave with periodic perturbation is  created by CW laser 1. The intensity and the phase of the pump and the sidebands are precisely controlled. The resulting 3-wave input is injected into a $L=18.28$ km long SMF-28 fiber (group velocity dispersion $\beta_{2}=-21 \times 10^{-27}\ \mathrm{s^{2}m^{-1}}$, nonlinear coefficient $\gamma=1.3 \times 10^{-3}\ \mathrm{W^{-1}m^{-1}}$). The loss is actively compensated by using a counter-propagating Raman pump emulating an almost fully transparent optical fiber \cite{Mussot2018}. Power and phase distributions of the pump and the first order sideband (signal) are obtained using a multi-heterodyning technique between the backscattered signal and the local oscillator \cite{Mussot2018}. 

In order to apply the theory in the previous section to optical fibers, the variables must be renormalized.
To this end, the dimensional distance $Z$, time $T$ (in the frame traveling at light group-velocity), and field $\Psi$ (with $|\Psi|^2$ giving directly the power in Watts) are obtained by the following rescaling
\begin{eqnarray}
& &Z = (z-z_0) L_{NL},\;\;T=t~{T_s},\;\;\Psi=\psi \sqrt{P_{P}}, \label{norm1} \\
& &L_{NL} = (\gamma P_{P})^{-1},\;\; {T_s}=\sqrt{|\beta_2|L_{NL}},\label{norm2}
\end{eqnarray}
where $L_{NL}$ is the characteristic nonlinear length scale associated with CW power $P_p$, and {$T_s$} is the relative temporal scale associated with dispersion.
Here $z_0$ is a suitable shift that accounts for the fact the input $Z=0$ corresponds to a point of weak modulation in the solution (whereas $z=0$ is the point of maximum amplification in the solution). For practical purposes, we can approximate $z_0 \approx \mathcal{Z}/4$, valid for weak enough input modulation.

\begin{figure}[htb]
\includegraphics[scale=0.50]{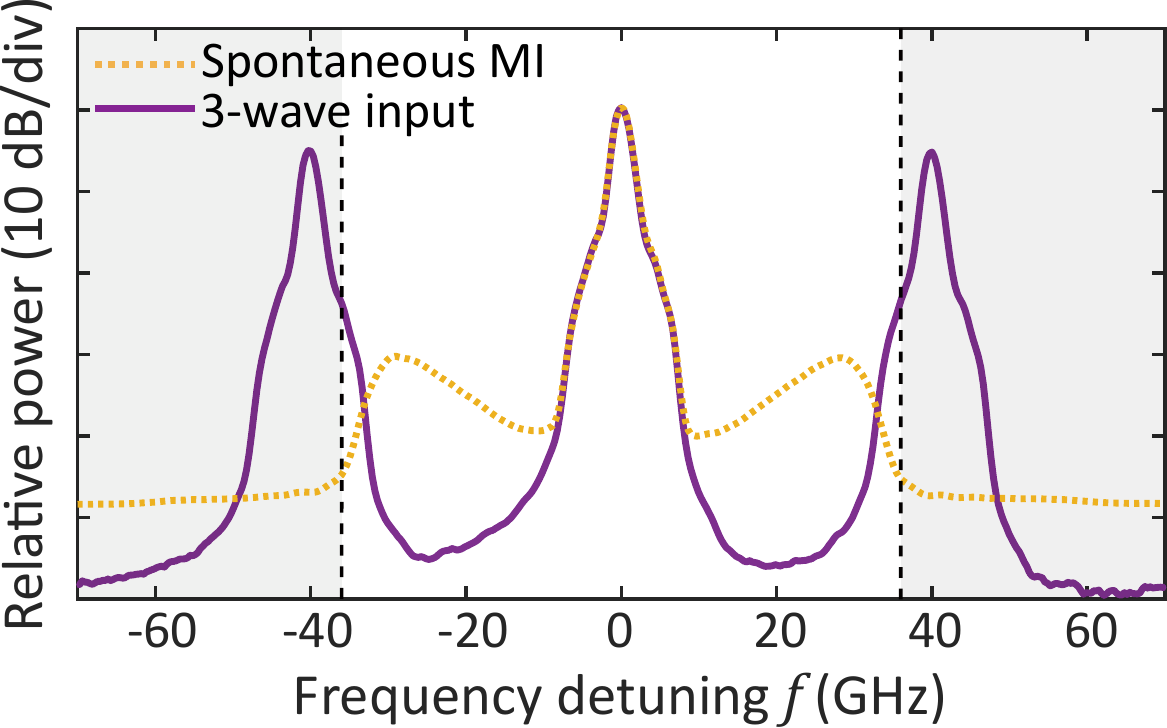}
\caption{Experimental 3-wave input spectrum (purple solid line) compared with the spontaneous MI spectrum (yellow dotted line). The two grey areas beyond dashed black lines mark the range of frequencies $f > f_{C}$, 
located outside of the conventional MI gain band ($f \le  f_{C}$).
}
\label{Fig1_opt}
\end{figure}

In this scaling, the MI cutoff frequency is 
 $f_{C}=2/(2\pi T_s)=1/(\pi \sqrt{\left| \beta_{2} \right|L_{NL}})$. 
The pump power $P_{P}$ in experiments is set to 180 mW leading t.o a cutoff frequency of the conventional MI gain band at  $f_{C}=33.6$ GHz. In all our experiments, the modulation frequency $f_{m}$ is located outside the MI gain band i.e. $f_{m} > f_{C}$. The intensity of the sidebands is set at 5.3 dB below the pump power. The experimental spectra of the 3-wave input and the spectrum of spontaneous MI, i.e. conventional MI gain band profile, are plotted in Fig.~\ref{Fig1_opt}. The initial relative phase between the pump and the signal is $-\frac{\pi}{2}$ in order to excite the A-type waves.

\begin{figure}[ht]
\includegraphics[scale=0.46]{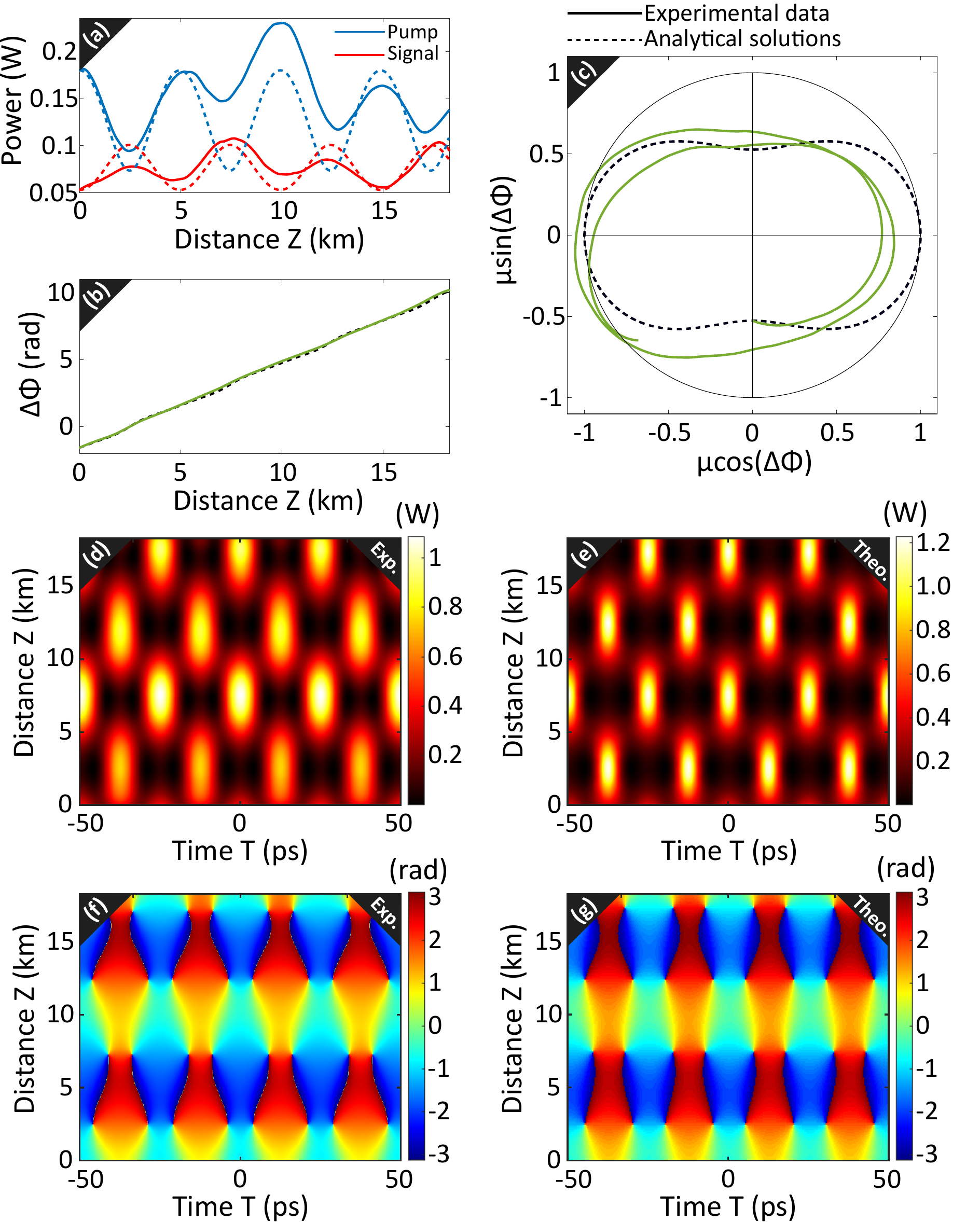}
\caption{Three waves evolution along the fiber when the sideband detuning $f_{m}=40$ GHz ($\omega_{m}=2.3809$) is outside the conventional MI band. (a) Evolution of the pump power (blue curves) and signal power (red curves). The solid curves in (a),(b) and (c) correspond to the experimental data while dashed curves are theoretical. (b) Relative phase vs distance. (c) The phase-plane representation of the evolution. (d) and (e) Spatio-temporal false-colour plots of the power profile. (f) and (g) The phase profiles. The panels (d) and (f) show the experimental data while (e) and (g) are theoretical. The following parameters have been used to prepare the initial conditions in the experiment and to plot the analytical solution: $\eta=1.2420$, $\rho=0.0317$, $\alpha_3=1.0$.}
\label{Fig2_opt}
\end{figure}

Experimental data for $f_{m}=40$ GHz and analytical solution for the same set of parameters are shown in Fig.~\ref{Fig2_opt}. 
Fig.~\ref{Fig2_opt}(a) displays the experimental power evolution of the pump (blue) and the signal (red). The corresponding theoretical prediction is shown by dashed curves. As expected, first, we observe amplification of the signal and depletion of the pump. The process reverses at around $2.5$ km when the maximum depletion of the pump is reached. The signal in experiment is amplified by $1.7$ dB between its initial value in $Z=0$ and its first maximum ($2.8$ dB gain for the corresponding analytical solution). The gain outside the conventional MI bandwidth is lower than the theoretical one shown in Fig~\ref{growth}(b) as the values of $\eta$ in experiments are higher (for $\omega=\omega_{m}$ and $\eta=0.05$ the theoretical gain is $5.2$ dB).

The first recurrence to the initial power profile occurs at $5$ km, and then successive cycles of growth and decay are repeated.
Overall, more than three periods of such oscillatory evolution can be seen in Fig.~\ref{Fig2_opt}(a). Deviation from perfect periodic dynamics is due to an imperfect compensation of loss by the active compensation system. The consequence of this inaccuracy is over amplification of the signal around 10 km mark.  
Fig.~\ref{Fig2_opt}(b) shows the nearly linear evolution of the pump-signal relative phase ($\Delta\Phi$) over the fiber length. The experimental curve fits the theory almost perfectly.
Importantly, the initial phase is recovered after two cycles of power evolution (around $Z=10$ km), whereas successive maximum amplification stages turn out to be mutually out of phase (sidebands shifted by $\pi$), which is a unique feature of A-type solutions \cite{PRA2020}.

It is also very convenient to illustrate the dynamics of the process in the phase space ($\mu\cos(\Delta\Phi)$, $\mu\sin(\Delta\Phi)$) where $\mu$ is the signal power normalised to its maximum value. Such trajectories are shown in Fig.~\ref{Fig2_opt}(c). The theoretical curve shown by the dashed line is strictly periodic and corresponds to the A-type solution. The quantitative agreement is also pretty good if we focus on the locations of the curve maxima. 
Figure~\ref{Fig2_opt}(c) also gives a pictorial view of the fact that the sidebands amplification is connected to the nonlinear deformation of the orbit with respect to circular orbits (characteristic of the purely linear limit $\omega \rightarrow \infty$, not shown). The net gain indeed arises, in the figure-of-eight-shaped orbit, from the ratio of the signal at the maximum elongation (horizontal axis, $\Delta\Phi=0,\pi$) and at the maximum orbital squeezing (input, $\Delta\Phi=-\pi/2$).
 
Figures \ref{Fig2_opt}(d) and (f) show the spatio-temporal evolutions of the power (Fig.~\ref{Fig2_opt}(d)) and phase (Fig.~\ref{Fig2_opt}(f)) of the electric field calculated from the inverse Fourier transform of the 3 main spectral components (see Figs.\ref{Fig2_opt}(a)-\ref{Fig2_opt}(c)). We used a procedure similar to that in Ref. \cite{naveau2019}. Characteristic chess-board-like pattern are obtained which is a clear signature of A-type solutions. The agreement with the analytical solution (Fig.~\ref{Fig2_opt}(e) and (g)) is very good. 
We notice, once again, that the input phase is recovered after two grow-decay cycles of power evolution, whereas successive maximal amplification profiles are shifted by half of the transverse period.

\begin{figure}[ht]
\includegraphics[scale=0.46]{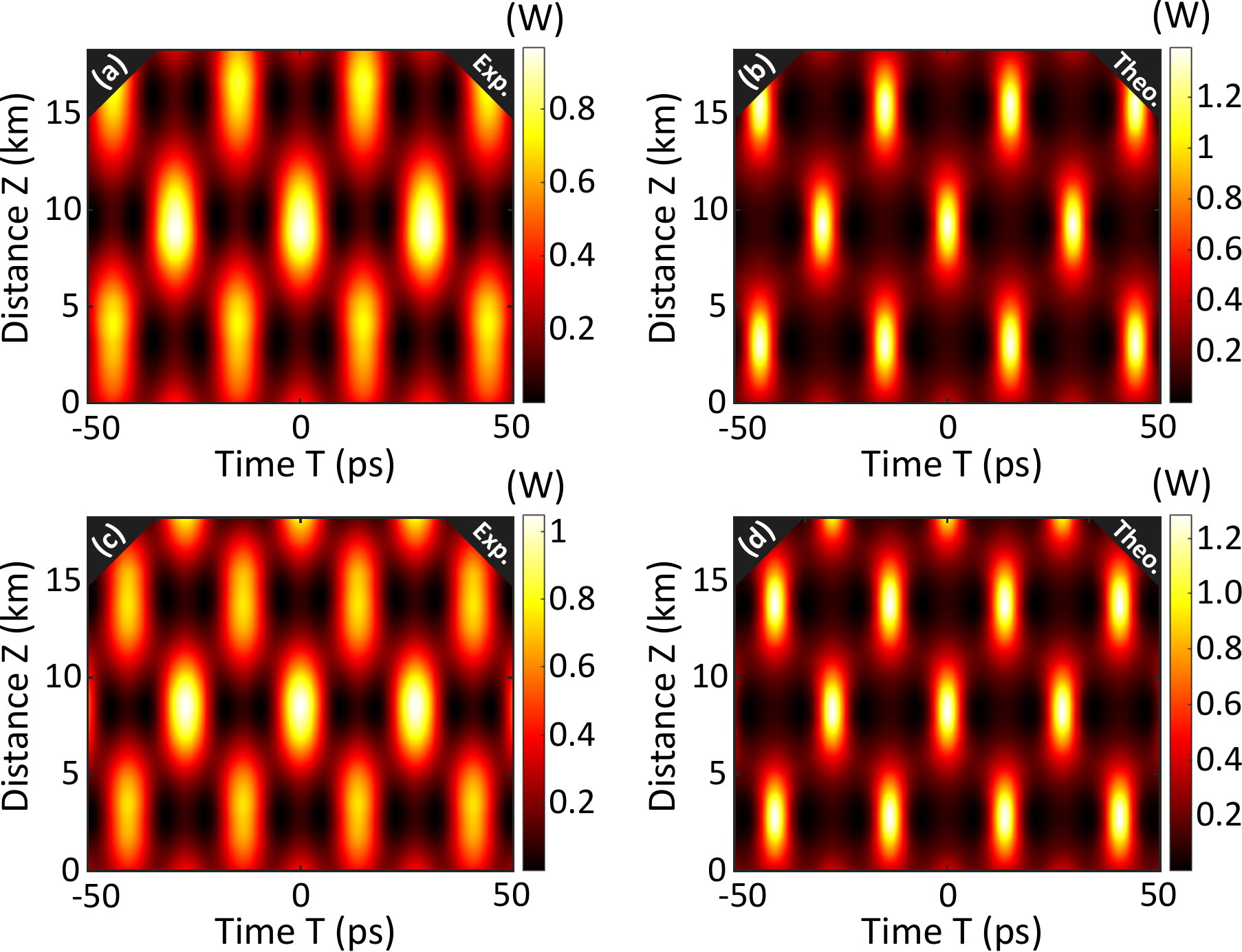}
\caption{Spatio-temporal power evolution for two other values of the pump-signal frequency shift. (a),(b) $f_{m}=34$ GHz ($\omega_{m}=2.0238$); (c),(d) $f_{m}=37$ GHz ($\omega_{m}=2.2023$). Left panels: experimental data, right panels: analytical solution. The following parameters have been used to prepare the initial conditions in the experiment and to plot the analytical solution: (a),(b) $\eta=1.0385$, $\rho=0.4275$, $\alpha_3=1.0$; and  (c),(d) $\eta=1.1407$, $\rho=0.2374$, $\alpha_3=1.0$.}
\label{Fig4_opt}
\end{figure}

Figure \ref{Fig4_opt} shows two additional spatio-temporal evolutions of the signal power from experimental measurements (left panels) and from analytics (right panels). 
The two signal frequency shifts ($f_m=$ 34 and 37 GHz respectively) are still outside of the MI band but located closer to the cutt-off frequency. Again, the chess-board like pattern of these plots confirms the A-type nature of these solutions. We can also notice that when approaching the cut-off frequency, the spatial periods (along $z$) increase, as can be seen from Figs.\ref{Fig2_opt} (d) and \ref{Fig4_opt} (a) and (c). Importantly, maximal wave amplitudes reached at the points of maximal compression are of the same order of magnitude for all cases shown in these figures.

\begin{figure}[ht]
\includegraphics[scale=0.46]{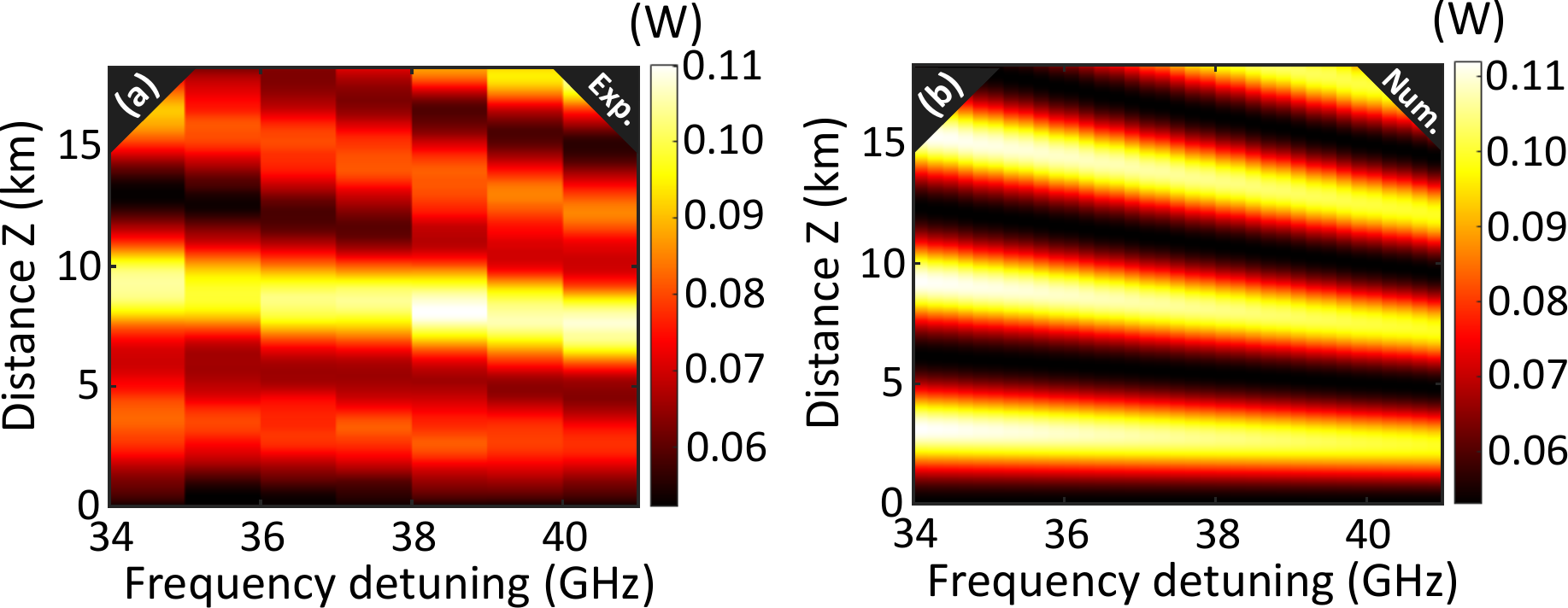}
\caption{2-D plot of the signal power as a function of distance $Z$ (vertical axis) and pump-signal frequency shift (horizontal axis). (a) experiment and (b) numerics.}
\label{Fig3_opt}
\end{figure}

As mentioned, the spatial (longitudinal) period depends on the shift between the modulation and the pump frequencies. When the modulation frequency is outside of the MI band, this period decreases with the modulation frequency moving away from the pump. Experimental verification of this behaviour is shown in Fig.~\ref{Fig3_opt}(a). Fig.~\ref{Fig3_opt}(b) shows the corresponding theoretical plot. 
While the frequency shift increases from 34 GHz to 41 GHz, the number of longitudinal periods along the same distance $\approx18$ km changes from 3 to 3.8. This means that each longitudinal period decreases from $\approx6$ km to $\approx4.73$ km. Agreement between the experimental data and the theory is also good as the two plots in Fig.~\ref{Fig3_opt}  demonstrate. Thus, our optical experiments confirm, clearly, the fact of existence of modulation instability outside of the conventional instability band. The measurements are in good agreement with the theoretical predictions based on the exact solutions of the NLSE.

\section{Water wave experiments}

\begin{figure}[ht]
\centering
\includegraphics[scale=0.52]{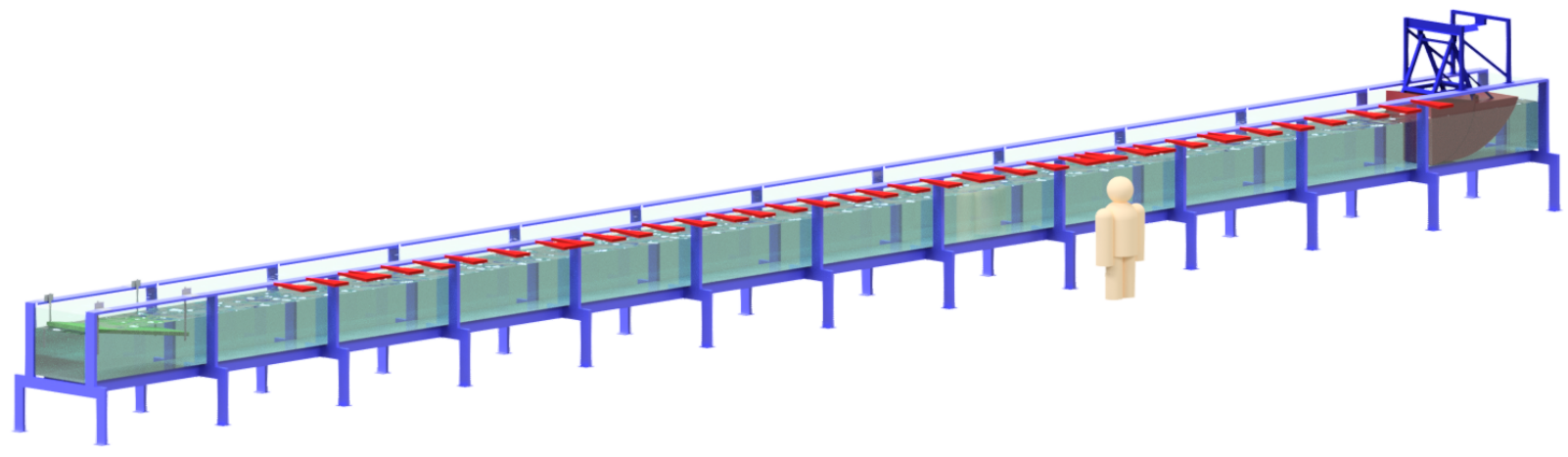}
\caption{Sketch of the L=30 meters long water tank at the University of Sydney. In red the locations of the gauges.}
\label{watertank}
\end{figure}

The hydrodynamic experiments have been performed in a uni-directional wave tank,
installed at the University of Sydney and shown in Fig.~\ref{watertank}. 
Its dimensions are $1$ m $\times$ 1 m $\times$ 30 m. The tank was filled with fresh water to the height of $0.7$ m in order to satisfy the deep water conditions for waves generated at the peak frequencies between 1.5 and 2.0 Hz. 
The piston wave maker with oscillation frequency range of $0.4<f<2$ Hz is installed at the right end of the tank. A wave absorbing beach was located at the opposite end to eliminate any influence from reflected waves. 
The piston is activated by an electric actuator, controlled by a pre-processed signal, which allows the seeding of a modulated surface elevation profile, according to mathematical expressions given above.

Eight wave gauges with a sampling rate of 32 Hz each are placed along the tank to collect the water wave elevation data. Due to repeatability of experiments, all eight wave gauges have been repositioned five times along the facility to ensure high resolution of the data acquisition both in time and in space. The gauges locations measured from the mean position of piston in these experiments are represented in red in Fig.~\ref{watertank}. This gave us sufficient resolution in $z$ for plotting the experimental patterns, as shown below.
The wave envelopes have been computed using the Hilbert transform while the spectral data have been calculated using the fast Fourier transform of the water surface elevation data.

Although the water wave envelope obeys the same dimensionless focusing NLSE as the normalized optical field in optical fiber, the spatial and time scales turn out to be extremely different.
We start from dimensional deep-water time-NLSE \cite{osborne2010nonlinear} characterized by the second-order dispersion coefficient $\beta_2=-2/g$ ($g=9.81$ m/s$^2$ is the gravitational acceleration) and the nonlinear coefficient $\gamma=-\kappa^3$ ($\gamma \beta_2 >0$, focusing regime), where $\kappa$ is the wavenumber of the carrier, with the carrier frequency fixed through the dispersion relation $\omega=\sqrt{g \kappa}$. 
In order to introduce a normalization akin to Eqs. (\ref{norm1}-\ref{norm2}), the dimensional distance along the tank $Z$, the dimensional time $T$, and the envelope of water wave elevation $\Psi$, can be expressed in terms of nonlinear length $L_{NL}$ and temporal scale $T_s$, fixed by the input envelope elevation $a$, 
as follows (see also \cite{ElKoussaifi2014})
\begin{eqnarray}
Z = (z-z_0)~L_{NL},\;\;T=t~T_s,\;\;\Psi=\psi~a, \label{normww1}\end{eqnarray}
where $L_{NL} = \frac{1}{\kappa^3 a^2}$, $T_s= \sqrt{\frac{2}{g \kappa^3 a^2}}$,
and $z_0$ is a suitable shift for which considerations analogous to those made in optics are still valid.
It is worth mentioning that in this case, the time $T$ is also measured in the frame moving with the group velocity $c_g=\frac{\omega}{2\kappa}$. Obviously, this scaling is not unique. An equivalent choice often employed in the case of water waves can be written in terms of the wavenumber $\kappa$ and the wave steepness $\varepsilon=a\kappa$: $Z=z/(\kappa \varepsilon^2)$, $T=\sqrt{2} t/(\omega {\varepsilon})$, $\Psi= \psi {\varepsilon}/\kappa$.

\begin{figure}[ht]
\includegraphics[scale=0.46]{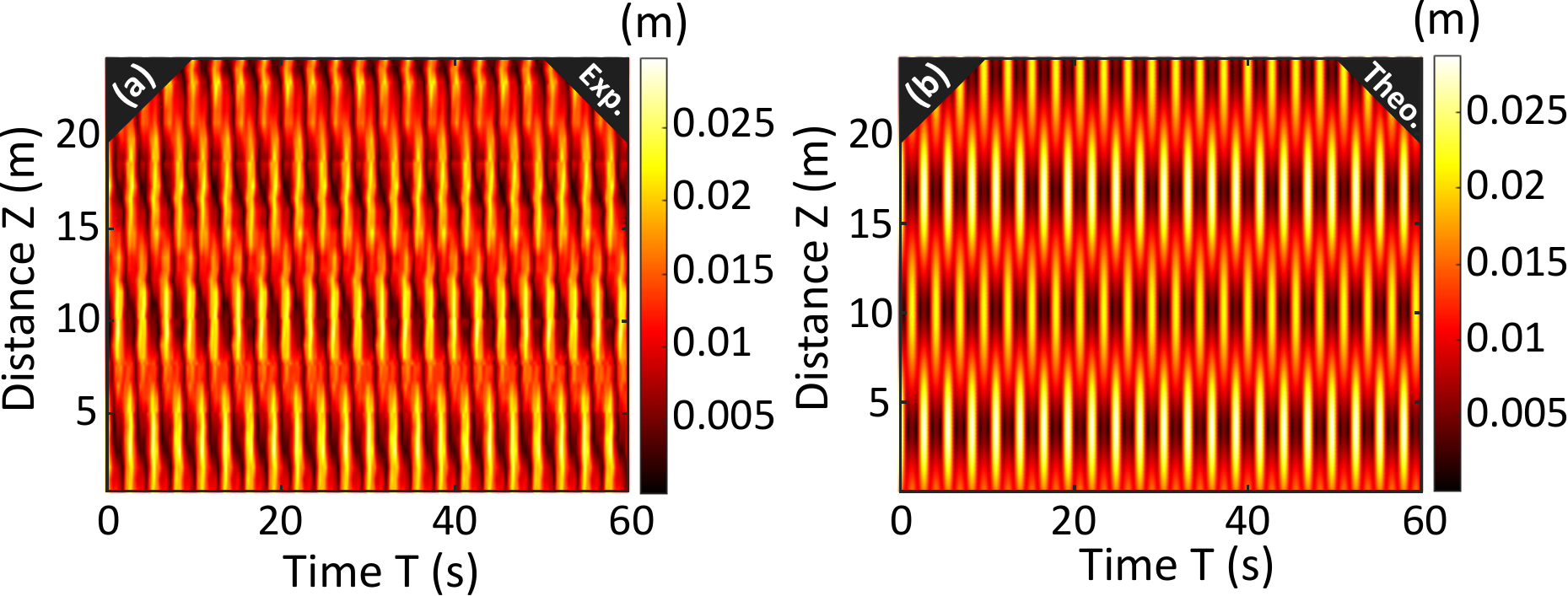}
\caption{(a) Experimental and (b) theoretical plots of spatio-temporal wave evolution that start with extraordinary modulation instability. The values of parameters in the NLSE solution used to prepare the initial conditions in the experiment are: $\eta=1.9$, $\rho=-0.9$ and $\alpha_3=1.0$. Wave amplitude $a=0.01$ m, the carrier wavenumber 
 $\kappa=10~\textnormal{m}^{-1}$, the corresponding wave steepness ${\varepsilon}=0.1$, and the modulation frequency $f_m=0.37$ Hz is well above the cut-off $f_C=0.22$ Hz.
}
\label{Fig7}
\end{figure}

\begin{figure}[ht]
\includegraphics[scale=0.46]{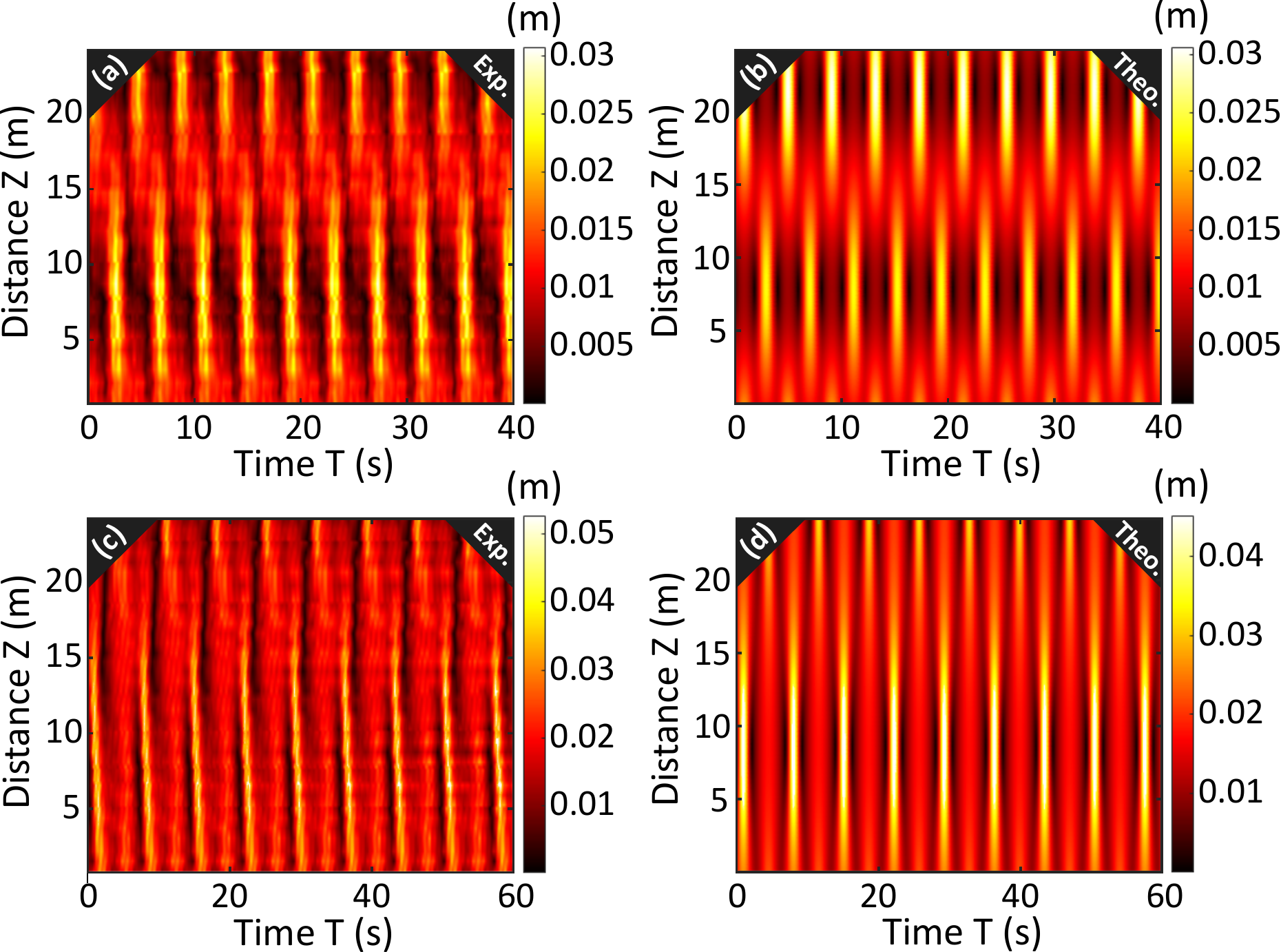}
\caption{(a,c) Experimental and (b,d) theoretical plots of spatio-temporal wave evolution that start with modulation instability.  The values of parameters in the NLSE solution used to prepare the initial conditions in the experiment are: (a,b) $\eta=1.03$, $\rho=0.355$, $\alpha_3=1.0$; and (c,d) $\eta=1.0$, $\rho=2.0$, $\alpha_3=1.0$. Wave amplitude $a=0.01$ m in each case. The wavenumber of the carrier is (a) 
 $\kappa=10~\operatorname{m}^{-1}$  and (c) $\kappa=8~\operatorname{m}^{-1}$. The corresponding wave steepness ${\varepsilon}=0.1$ with cut-off frequency $f_C=0.22$ Hz in (a) and $a\kappa=0.08$ with cut-off frequency $f_C=0.15$ in (c). 
The modulation frequency is $f_m=0.25$ Hz in (a) and $f_m=0.16$ Hz in (c).
}
\label{Fig8}
\end{figure}

\begin{figure}[ht]
\includegraphics[scale=0.46]{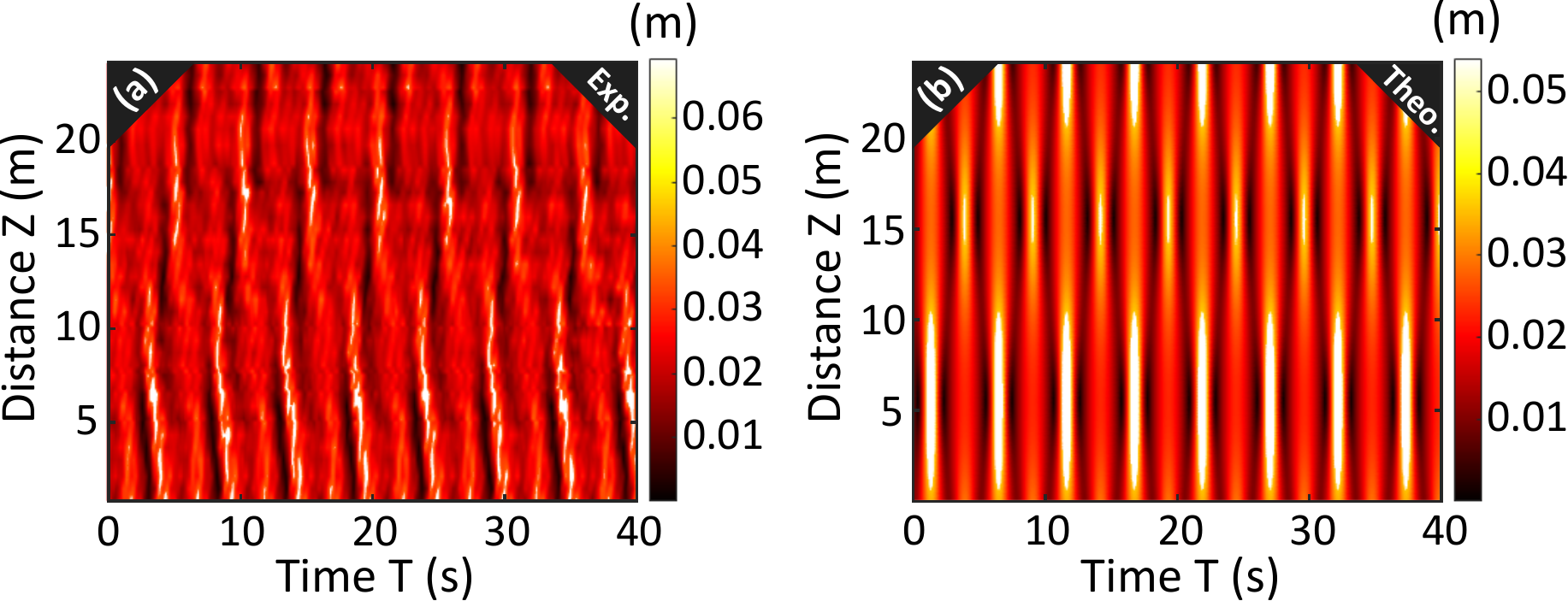}
\caption{(a) Experimental and (b) theoretical plots of spatio-temporal wave evolution that start with modulation instability. The values of parameters in the NLSE solution used to prepare the initial conditions in the experiment are: $\eta=1.9$, $\rho=2.9$ and $\alpha_3=1.0$. Wave amplitude $a=0.01$ m. The wavenumber of the carrier in (a) is 
 $\kappa=8~\operatorname{m}^{-1}$, consequently the wave steepness $\varepsilon=0.08$. The modulation frequency here $f_m=0.2$ Hz is once again above the cut-off frequency $f_C=0.15$ Hz. Modulation instability develops but wave breaking prevents the recurrence back to initial conditions.
}
\label{Fig9}
\end{figure}

Operating with the scaling in Eq. (\ref{normww1}), we obtained the theoretical spatio-temporal patterns shown in Figs.~\ref{Fig7}-\ref{Fig9} [see right panels (b)],
which we compare with experimental data [left panels (a) in the same figures]. The choice of the parameters of the NLSE solution used for generating these patterns is given in the figure captions. In our water waves experiment, typically, $L_{NL} \approx 10$ m and ${T_s} \approx 1.4$ s, compared with $L_{NL} \approx 4$ km and ${T_s} \approx 10$ ps of the optical experiment. Accordingly, also the MI cut-off frequency, which reads, in this case, $f_C=\frac{1}{\pi} \sqrt{\frac{g \kappa^3 a^2}{2} }$, turns out to be several orders of magnitude lower than the one in optics. For instance, with $\kappa=10$ m$^{-1}$ and $a=0.01$ m (case of Fig.~\ref{Fig7}), we obtain $f_C=0.22$ Hz. The envelope evolution as predicted by theory takes into account the second-order Stokes correction to the water surface elevation \cite{dudley2019rogue}.\\

These spatio-temporal patterns are very similar to those obtained in optical experiments. Remarkably, our maxima (two periods) have been achieved within the length of the tank as can be seen from Fig.~\ref{Fig7}. The resulting chessboard structure of this pattern corresponding to the A-type doubly periodic wave is also clearly seen. Three recurrences to a nearly constant amplitude wave are clearly visible despite unavoidable dissipation elements, always present when performing laboratory experiments. Note that for the given carrier wave parameters, it would not be possible to observe more than one cycle of MI-growth-decay or AB within the given effective propagation distance of 25 m.
 
In order to reaffirm the observation, two more examples of the spatio-temporal pattern are shown in Fig.~\ref{Fig8}. 
These plots contain less then one period of evolution that includes one full recurrence to initial conditions at around $15$ meters mark in (a) and around $19$ meters mark in (c). In each case, the carrier steepness $\varepsilon$ has been adjusted to be just below the threshold of wave breaking. The latter happens due to the excessive wave amplitude amplification. 

One essential difference of experimental patterns in Figs.~\ref{Fig7}(a) and \ref{Fig8}(a)
from the optical ones is slightly tilted vertical stripes. The reason is the asymmetry of the water wave profiles, which is the result of significant breather amplification of a factor of three and above. The consequence is the nonlinear Stokes contributions that are always present in water waves at these scales \cite{dysthe1979note,Waseda}. Despite these deviations, the patterns in Figs.~\ref{Fig7}(a) and \ref{Fig8}(a) clearly confirm the presence of the modulation instability and its nonlinear evolution beyond the standard unstable frequencies of MI in the BF theory.

Generally, when increasing the amplification factor of the breather, the steepness has to be decreased in order to avoid wave breaking. The latter violates the condition of the flow to be irrotational and thus, prohibits applicability of the Euler equations and subsequently, the  validity of the NLSE \cite{mei1983applied}. Indeed, when this threshold of wave breaking is exceeded, spilling as well as recurrent breaking occurs and the pattern changes significantly and does not follow the theoretical NLSE predictions. One example is shown in Fig.~\ref{Fig9}.
Here, the value of the breather parameter $\rho$ is increased in comparison to the previous cases. Modulation instability still develops but there is no obvious recurrence back to initial conditions as can be seen from Fig.~\ref{Fig9}(a).

\section{Conclusions}

In conclusion, we have experimentally confirmed that modulation instability is more complex phenomenon than the one predicted by the simplified linear stability analysis. The most striking difference that the more accurate nonlinear analysis using exact breather framework reveals is the fact that periodically perturbed continuous waves develop the growth of perturbation not only within the standard modulation instability band described by the BF and BT theories but also outside of it. To be more accurate, the frequency range of unstable growth of the perturbation extends beyond the standard MI threshold.

Another dramatic difference from the standard theory can be seen when observing the nonlinear stage of MI. The subsequent evolution beyond the initial growth creates a specific chess-board like periodic spatio-temporal pattern of wave propagation. Temporal maxima of the generated pulse trains change position every half period of spatial evolution. The effect tightly related to this phenomenon is the fact that the frequency of the pulse train at the point of maximum compression is half of the frequency of initial modulation. Such phenomenon may find applications in frequency comb devices facilitating the atomic clock synchronisation when the frequencies differ by an octave \cite{Newbury2011}.

Having these unusual features revealed in nonlinear analysis, we can call the effect `extended' or `extraordinary' modulation instability. Importantly, we were able to track and confirm this `extraordinary' modulation instability in two different physical media, namely, in optics and in hydrodynamics, proving the interdisciplinary significance of the extended MI. In fact, these are the two areas of physics where the wavelength differs by four orders of magnitude, and the modulation frequencies by ten orders of magnitude.  
This twofold confirmation of the effect shows that it is ubiquitous and does not depend on the scale of the physical system that we operate with. The effects should be also observable 
in other areas of physics such as astrophysics \cite{Astro}, plasma \cite{Ichikawa73,Thornhill1978plasma}, meta-materials \cite{Boardman2020} and in Bose-Einstein condensate \cite{Hulet2017,Everitt2017BECexp}.
We envisage that the new phenomenon can be useful in applications such as generation of optical frequency combs and pulse trains with prescribed parameters: periods, amplitudes and duty cycles. Moreover, we anticipate novel modelling approaches for extreme events in nonlinear dispersive media.

\section{Acknowledgments}
Agence Nationale de la Recherche (Programme Investissements
d’Avenir); Ministry of Higher Education and Research;
Hauts de France Council; European Regional Development
Fund (Photonics for Society P4S, FUHNKC, EXAT).

\end{document}